# A FRAMEWORK FOR COMPUTING TRANSPORT PROPERTIES OF CARBON NANOTUBE-BASED CONDUCTANCE BIOCHEMICAL SENSORS

*C. Roman, F. Ciontu, B. Courtois*

Tima Laboratory INPG/UJF/CNRS, Grenoble, France

## ABSTRACT

In this paper we present a framework for fast quantum conductance calculations of carbon nanotube-based sensing devices targeting aromatic amino acids within a tight binding approximation. The method begins by a novel parameterization procedure based on isospectral matrix flows. With the properly parameterized Hamiltonian we employ a linearly scaling algorithm to compute the quantum conductance in the coherent transport regime. A few conclusions are presented regarding the suitability of carbon nanotubes in aromatic amino acid detection.

## 1. INTRODUCTION

The development of biology seems to increasingly depend on the availability of selective biochemical sensors capable to determine for instance the amino acid composition of a protein. That alone is often enough to identify a protein [1] or even predict its secondary structure [2,3]. However, the required sensitivity and dynamic range rule out most of the potential sensing mechanisms. As carbon nanotubes emerge as a very promising alternative to now standard conductance thin-film sensors we have assumed the task of studying whether sensing amino acids is possible via this paradigm.

Carbon nanotube-based chemical sensors have been experimentally demonstrated for $NO_2$, $NH_3$ [4], $H_2$ [5] and $O_2$ [6]. In this paper we will focus on similar devices that respond to amino acid adsorption with a change in their conductance. The study will be limited to zwitterion aromatic Histidine (HIS), Phenylalanine (PHE), Tryptophan (TRP) and Tyrosine (TYR) amino acids, binding through π stacking onto single-walled carbon nanotubes.

A direct *ab initio* simulation of a carbon nanotube conductance sensor is prohibitively difficult, primarily because of the huge number of atoms involved that rules out direct diagonalization approaches. In a previous paper [7, 8] we described how self-consistency can be avoided altogether by properly choosing a considerably smaller reference system, namely each of the four amino acids on top of a flat graphene sheet. There is also another reason in doing so, since studying the density of states of the graphene sheet with adsorbed amino acids could reveal if, at least in theory, the conductance of a nanotube would be perturbed by the presence of these molecules.

## 2. AB INITIO ELECTRONIC STRUCTURE

A more detailed description of the *ab initio* calculations performed on these systems is contained elsewhere [7]. Here we focus only on those results considered relevant for the logic of our explanation.

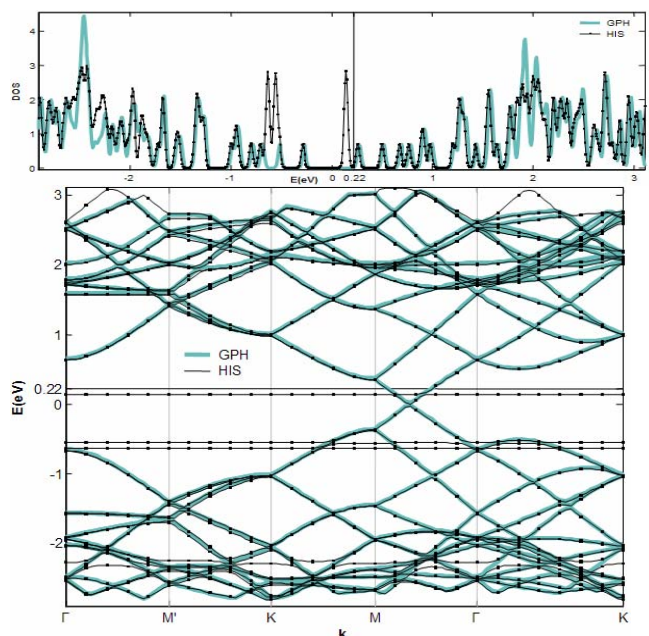

Figure 1: Total density of states and band structures for GPH+HIS. The light curves correspond to pristine GPH

Figure 1 displays the total density of states (TDOS) and band structure as obtained with SIESTA [9], in which the reference, pristine graphene (GPH) properties are plotted together with the same properties but for the GPH+HIS system. Typically the physisorption is found to shift the Fermi level by 200 meV, and introduces dispersionless bands close to $E_F$, whose positions depend on the amino acid, even though they were all initiated by the same α-carboxyl group. The existence of physisorption induced states close to $E_F$ is considered a necessary condition for carbon nanotubes to be susceptible of detecting aromatic zwitterion amino acids.

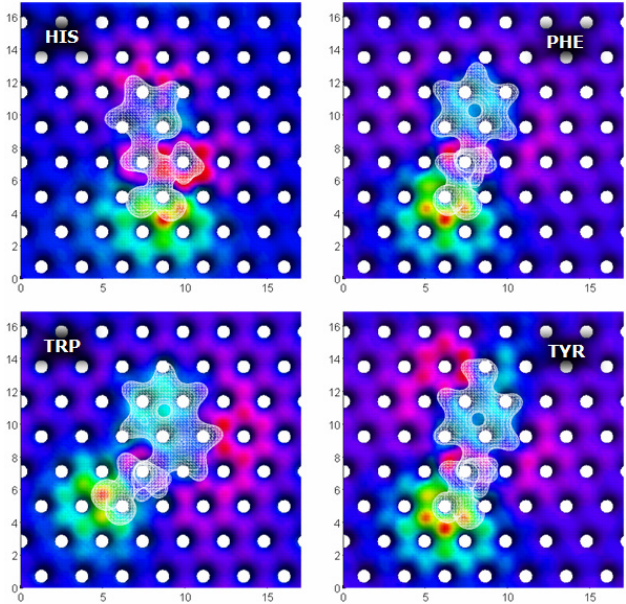

Figure 2. Electron density of the graphene plus amino acids, mapped on an isocharge density surface of the pristine graphene

A mapping of the charge density of graphene plus amino acids on an isocharge density surface of the pristine graphene, gives a qualitative measure of the physisorption-induced charge redistribution. As revealed in Figure 2 the charge perturbation is strongly localized, not surprisingly around the same α-carboxyl group. The spatial confinement of the perturbation establishes the validity of approximating amino acids-on-nanotube Hamiltonian matrix elements by their equivalents within a reference, amino acid on graphene system.

## 3. HAMILTONIAN MODEL REDUCTION AND RENORMALIZATION

Our aim is to compute the quantum conductance of a large atom number system. No matter how close to order-N a conductance calculation method is, its scaling pre-factor is inevitably proportional to the third power of the mean number of basis orbitals per atom (MNOA). For instance using a Double Zeta (DZ) basis in SIESTA for the Histidine on graphene system yields 1002 basis orbitals, provided that the number of atoms is only 132, i.e. approximately 7.6 (760%) MNOA. This raises a major obstacle in the way of conductance calculations since a realistic CNT sensor would have at least in the order of $10^4$ atoms and consequently $10^5$ basis elements with respect to the reference system that involved only about 1000 basis elements.

Naturally the first thing one tries to solve this problem is to come up with a reduced order Hamiltonian model that has a certain spectral accuracy but only within a certain energy range. This considerations are at the base of top-down tight binding (TB) Hamiltonian parameterization, successfully reproducing calculations of computationally intensive ab initio methods, with just a modest mean number of basis orbitals per atom, examples including pristine graphite and carbon nanotubes [10], boron/nitrogen doped SWNTs [11], any many others. However the systems studied in this paper are profoundly asymmetric and this seriously limits a straightforward application of the same model reduction schemes that were suitable for structures enjoying high symmetry. Asymmetry increases considerably the number of parameters in top-down schemes slowing down or even forbidding the parameter space exploration. Furthermore the optimization would most probably stuck in a local minimum yielding decent parameters for the reference system, but, since there is no formal framework in guiding the search, there is no guarantee that the same set of parameters can be extrapolated to other systems.

In this view we have developed a completely automated method of minimizing the number of basis elements per atom and at the same time preserving the spectral accuracy around the Fermi level which determines entirely the charge transport properties of a system. This method starts from the self-consistent Hamiltonian and overlap set of matrices and produces, in a bottom-up manner, a new set via isospectral congruence transformations.

In selecting which basis orbitals to reduce, we exploited the spectral information stored in the projected density of states (PDOS) on each basis element. This was multiplied by an energy filtering function, here a Gaussian distribution centered at $E_F$ and of dispersion $(\mu_2-\mu_1)/3$ and integrated over the energy range of interest $[\mu_1, \mu_2]$. In Equation 1, $n_\lambda(E)$ is the PDOS as provided by SIESTA that is already integrated over the first Brillouin Zone (BZ).

$$n_\lambda = \int_{\mu_1}^{\mu_2} dE \cdot \exp\left(-\frac{9}{2}\frac{(E-E_F)^2}{(\mu_2-\mu_1)^2}\right) \cdot n_\lambda(E) \qquad (1)$$

Next, those basis elements $\{|\lambda\rangle\}$, for which $n_\lambda$ fell bellow a given threshold ($\varepsilon_n$) were eliminated from matrices $H_0$ and $S_0$ by suppressing corresponding lines and columns. In the case of Histidine having set $\mu_i$ to $\pm 1eV$ and $\varepsilon_n$ to 0.5% reduced the number

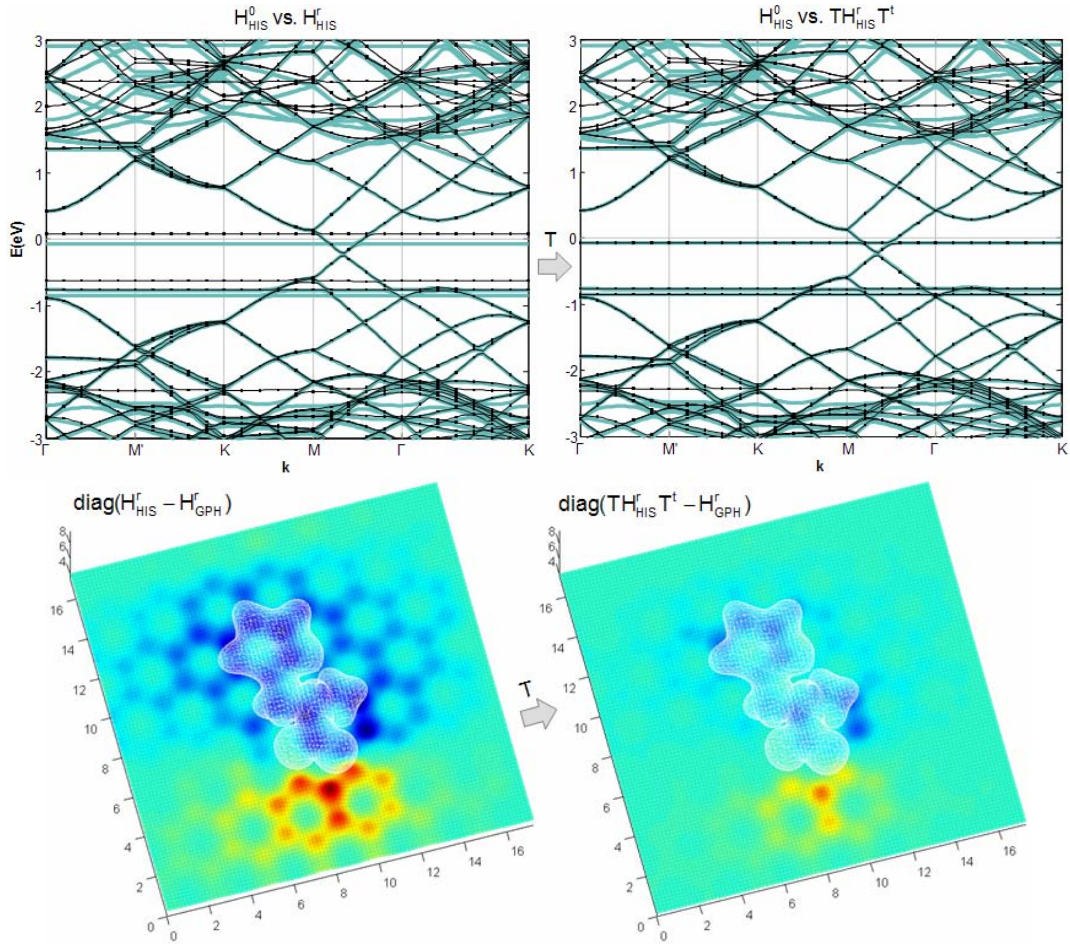

Figure 3: (Top) HIS band structure after orbital elimination (left) and after renormalization (right). (Bottom) Diagonal Hamiltonian elements represented in real space before (left) and after renormalization (right)

of orbitals from 1002 to 141, which is very close to 132 that represents the total number of atoms. This reduction alone would yield a 400 times speed-up in the conductance calculation procedure.

### 3.1. Isospectral matrix flows

The sub-space projection described in the previous section has one obvious problem in practice. The elimination of atomic orbitals from the basis set perturbs, the band structures as can be observed for instance in Figure 3 top-left.

A second problem, which has more to do with the limited size of the GPH super cell than with the orbital elimination procedure, made us consider renormalizing the Hamiltonian and overlap matrices ($\mathbf{H}$, $\mathbf{S}$), i.e. modifying their elements so as to simultaneously satisfy given spectral and structural constraints.

Our renormalization procedure is an adaptation of Chu's least squares approximation of symmetric-definite pencils subject to generalized spectral constraints [12] to which we refer the reader for a rigorous introduction to the mathematical concepts. In simple terms the procedure consists in finding a pair of matrices ($\mathbf{H}$, $\mathbf{S}$) that yield the same eigenvalues as some pair ($\mathbf{H}_0$, $\mathbf{S}_0$) and are as close as possible, element-wise, to some other pair ($\mathbf{H}_\perp$, $\mathbf{S}_\perp$).

As the set:

$$\mathcal{M}(\mathbf{H}_0, \mathbf{S}_0) := \left\{ \left( \mathbf{TH}_0\mathbf{T}^t, \mathbf{TS}_0\mathbf{T}^t \right) \in \mathbb{R}^{n \times n} \times \mathbb{R}^{n \times n} / \det(\mathbf{T}) \neq 0 \right\}$$

consists of *all* symmetric definite pairs having the same eigenvalues with ($\mathbf{H}_0$, $\mathbf{S}_0$), the problem reduces to finding a congruence transformation matrix $\mathbf{T}$ such that ($\mathbf{H}$, $\mathbf{S}$) ≡ ($\mathbf{TH}_0\mathbf{T}^t$, $\mathbf{TS}_0\mathbf{T}^t$) optimally approximates ($\mathbf{H}_\perp$, $\mathbf{S}_\perp$). Formally, this is equivalent to finding the minimum of Equation 2, where ∘ is the Hadamard matrix product and $\|\mathbf{X}\|_F^2$ the Frobenius matrix norm. The two weighting matrices ($\mathbf{W}_\mathbf{H}$, $\mathbf{W}_\mathbf{S}$) represent the only distinction between our renormalization procedure and Chu's theory [12]. They allow in a straightforward manner to increase, decrease or even cancel any individual matrix element of **H** or **S**.

$$F(\mathbf{T}) := \frac{1}{2}\left(\|\mathbf{W_H} \circ (\mathbf{H} - \mathbf{H_\perp})\|_F^2 + \|\mathbf{W_S} \circ (\mathbf{S} - \mathbf{S_\perp})\|_F^2\right) \quad (2)$$

One major feature of isospectral flows is that the gradient of $F(\mathbf{T})$ is analytically computable; in our case

$$\nabla F(\mathbf{T}) = 2\left(\left[(\mathbf{W_H})_\circ^2 \circ (\mathbf{H} - \mathbf{H_\perp})\right]\mathbf{TH_0} + \left[(\mathbf{W_S})_\circ^2 \circ (\mathbf{S} - \mathbf{S_\perp})\right]\mathbf{TS_0}\right) \quad (3)$$

which makes it possible to quickly set-up a steepest descent flow in order to find the minimizer of Equation 2, i.e. the solution to our problem.

$$\dot{\mathbf{T}}(t) := -\nabla F(\mathbf{T}(t)) \quad (4)$$

We have applied this method to GPH+HIS system, whose properties before the renormalization process are found in the left column of Figure 3, where the two problems mentioned at the beginning of this sub-section are clearly visible; some bands are perturbed and the charge redistribution following physisorption extends throughout the unit cell. Due to the renormalization procedure's flexibility we were able to address simultaneously the two, apparently different, problems.

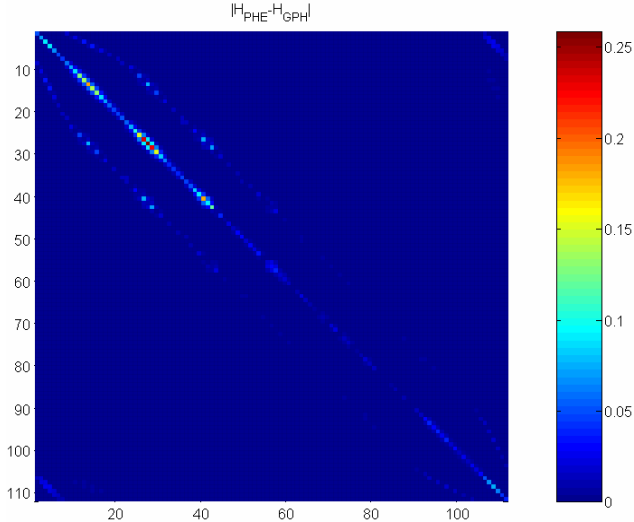

Figure 4. Absolute Hamiltonian matrix element difference between the graphitic parts of GPH+PHE and pristine GPH.

Figure 4 reveals another interesting aspect. As it represents the absolute Hamiltonian matrix element difference between the graphitic parts of GPH+PHE and pristine GPH, it shows clearly the localization of the charge perturbation due to physisorption and at the same time explains the 200 meV difference in the Fermi levels of the two systems.

## 4. EFFICIENT QUANTUM CONDUCTANCE CALCULATION

In this paper we consider only elastic transport within mean field theories like DFT, HF of TB, in which case Landauer-Büttiker like formulas are typically employed for computing the currents through molecular structures of the kind sketched in Figure 5.

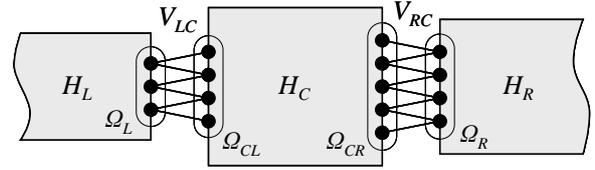

Figure 5. A generic two terminal molecular device and its real-space partitioning.

Within an atomic-like basis set, like in our case, the system's Hamiltonian and overlap matrices can easily be partitioned in the form

$$H = \begin{pmatrix} H_L & & H_{LC} \\ & H_R & H_{RC} \\ H_{LC}^\dagger & H_{RC}^\dagger & H_C \end{pmatrix}; \quad S = \begin{pmatrix} S_L & & S_{LC} \\ & S_R & S_{RC} \\ S_{LC}^\dagger & S_{RC}^\dagger & S_C \end{pmatrix} \quad (5)$$

A central quantity in the Landauer-Büttiker formalism is the Green's function or resolvent

$$G(z) = [zS - H]^{-1} = \begin{pmatrix} G_L & G_{LR} & G_{LC} \\ G_{RL} & G_R & G_{RC} \\ G_{CL} & G_{CR} & G_C \end{pmatrix} \quad (6)$$

With these notations the Green's function of the central region can easily be obtained using

$$G_C(z) = [zS_C - H_C - \Sigma_L(z) - \Sigma_R(z)]^{-1} \quad (7)$$

where the lead self-energies $\Sigma_{L(R)}(z)$ are given by

$$\Sigma_{L(R)}(z) = V_{L(R)C}^\dagger G_{L(R)}^0(z) V_{L(R)C}$$
$$G_{L(R)}^0(z) = [zS_{L(R)} - H_{L(R)}]^{-1}, \quad V_{L(R)C} = zS_{L(R)C} - H_{L(R)C} \quad (8)$$

The Landauer-Büttiker formula that gives the conductance at energy E reads:

$$\mathcal{G}(E) = \frac{2e^2}{h} \text{Tr}\left[ \Gamma_L(E) G_C^r(E) \Gamma_R(E) G_C^a(E) \right] \quad (9)$$

The retarded and advanced quantities employed in Equation 9 follow the general convention $G^{r(a)}(E) = G(z = E \pm i0^+)$. The leads are included as boundary conditions into the central region via $\Gamma_{L(R)}(E)$ that are defined as $\Gamma_{L(R)}(E) = i\left[ \Sigma_{L(R)}^r(E) - \Sigma_{L(R)}^{r\dagger}(E) \right]$.

Efficient conductance calculation schemes exploit the zero elements of the matrices involved in constituent equations. From Figure 5 it is easy to observe that $V_{L(R)C}$ has non-zero matrix elements only when linking orbitals found in the boundary regions $\Omega_{L(R)}$ in the leads and $\Omega_{CL(R)}$ in the central region respectively, i.e.

$$\left\langle V_{L(R)C} \right\rangle_{\mu\nu} = \delta_{\mu\Omega_{L(R)}} \delta_{\nu\Omega_{CL(R)}} \left\langle V_{L(R)C} \right\rangle_{\mu\nu} \quad (10)$$

In our notation, $\left\langle V_{L(R)C} \right\rangle_{\mu\nu} = \left\langle \mu | V_{L(R)C} | \nu \right\rangle$. Replacing Equation 10 into (8) yields:

$$\left\langle \Sigma_{L(R)}(z) \right\rangle_{\mu\mu'} = \sum_{\nu\nu'} \left\langle V_{L(R)C}^\dagger \right\rangle_{\mu\nu} \left\langle G_{L(R)}^0(z) \right\rangle_{\nu\nu'} \left\langle V_{L(R)C} \right\rangle_{\nu'\mu'} =$$
$$= \delta_{\mu\Omega_{CL(R)}} \delta_{\mu'\Omega_{CL(R)}} \sum_{\nu\nu' \in \Omega_{L(R)}} \left\langle V_{L(R)C}^\dagger \right\rangle_{\mu\nu} \left\langle g_{L(R)}^0 \right\rangle_{\nu\nu'} \left\langle V_{L(R)C} \right\rangle_{\nu'\mu'} \quad (11)$$

which means that $\left\langle \Sigma_{L(R)}(z) \right\rangle_{\mu\mu'}$ will be non-zero only when orbitals $\mu, \mu' \in \Omega_{CL(R)}$ and moreover only $\left\langle G_{L(R)}^0(z) \right\rangle_{\nu\nu'}$ with $\nu, \nu' \in \Omega_{L(R)}$ need to be calculated. The last term represents exactly the surface Green's functions $\left\langle g_{L(R)}^0(z) \right\rangle_{\nu\nu'}$ for which efficient calculation schemes exist already [13].

Consequently

$$\left\langle \Gamma_{L(R)}(E) \right\rangle_{\mu\mu'} = \delta_{\mu\Omega_{CL(R)}} \delta_{\mu'\Omega_{CL(R)}} \left\langle \Gamma_{L(R)}(E) \right\rangle_{\mu\mu'} \quad (12)$$

and Equation 9 reduces to:

$$\mathcal{G}(E) = \frac{2e^2}{h} \sum_{\substack{\mu\mu' \in \Omega_{CL} \\ \nu\nu' \in \Omega_{CR}}} \left\langle \Gamma_L \right\rangle_{\mu\mu'} \left\langle G_C^r \right\rangle_{\mu'\nu'} \left\langle \Gamma_R \right\rangle_{\nu'\nu} \left\langle G_C^r \right\rangle_{\mu\nu}^* \quad (13)$$

Again the last Equation shows that knowing the retarded Green's functions $\left\langle G_C^r \right\rangle_{\mu\nu}$ linking boundary orbitals $\mu \in \Omega_{CL}, \nu \in \Omega_{CR}$ suffice to compute $\mathcal{G}(E)$.

Summarizing, the conductance of a two terminal device is completely determined by the knowledge of the lead surface Green's functions $\left\langle g_{L(R)}^0(z) \right\rangle_{\nu\nu'}$ with $\nu, \nu' \in \Omega_{L(R)}$ and the left to right Green's functions of the central region $\left\langle G_C^r \right\rangle_{\mu\nu}$ with $\mu \in \Omega_{CL}, \nu \in \Omega_{CR}$.

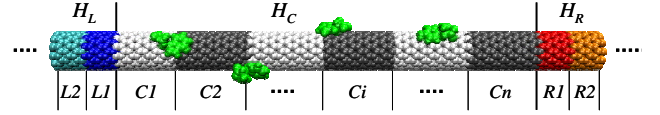

Figure 6. Hamiltonian of the carbon nanotube toy-sensor and its real-space partitioning.

Realistic sensor geometries including large metallic contacts bridged by long nanotubes remains currently a formidable task. In this paper we opted for a simplified toy sensor as in Figure 6 that has a central region $H_C$, coupled to two semi-infinite leads made of same-chirality nanotubes.

The lead surface Green's functions $\left\langle g_{L(R)}^0(z) \right\rangle_{\nu\nu'}$, which is the first quantity of interest can easily be obtained iteratively from the relation [13] (See Figure 6):

$$\mathbf{g}_{L(R)}^0(z) = \left[ z\mathbf{S}_{L(R)1} - \mathbf{H}_{L(R)1} - \mathbf{V}_{L(R)12}^\dagger \mathbf{g}_{L(R)}^0(z) \mathbf{V}_{L(R)12} \right]^{-1} \quad (14)$$

However the most difficult part is to compute $\left\langle G_C^r \right\rangle_{\mu\nu}$. Recently a promising recursion method was developed for the precise purpose of computing this term [15]. The method involves a two-sided Lanczos process for non-hermitian matrices and consequently its numerical stability must be considered carefully. Moreover a recursion terminator is not as easy to find as is for the hermitian case.

In this light we opted for a recursive method that is exact and also order-N at least for 1D systems like carbon nanotubes are [14]. This method exploits the block tri-diagonal structure of $\mathbf{H_C}$ and $\mathbf{S_C}$ when partitioned according to Figure 6.

The algorithm starts from

$$\begin{cases} \mathbf{A_0} = \mathbf{K_{11}} \\ \mathbf{B_0} = \mathbf{I} \end{cases} \quad (15)$$

where by notation $\mathbf{K_{ij}} = z\mathbf{S_{Cij}} - \mathbf{H_{Cij}}$ and computes recursively the quantities

$$\begin{cases} \mathbf{A_i} = \mathbf{K_{i+1i+1}} - \mathbf{K_{i+1i}} \mathbf{A_{i-1}^{-1}} \mathbf{K_{ii+1}} \\ \mathbf{B_i} = -\mathbf{B_{i-1}} \mathbf{A_{i-1}^{-1}} \mathbf{K_{ii+1}} \end{cases} \quad (16)$$

Then, the left to right retarded Green's function is simply

$$\mathbf{G_{C1n}} = \mathbf{B_{n-1} A_{n-1}^{-1}} \qquad (17)$$

The renormalized Hamiltonian and overlap matrices as obtained through the method described in Section 3 were used then for obtaining the matrix elements of a semiconducting (11, 0) nanotube and its variants with adsorbed amino acids.

The active region C of the toy sensor as depicted in Figure 6 measured approximately 15nm. Leads were simulated by ideal semi-infinite (11, 0) tubes. Histidine molecules were added randomly along the tube, but a minimum spacing was imposed between two neighboring amino acids.

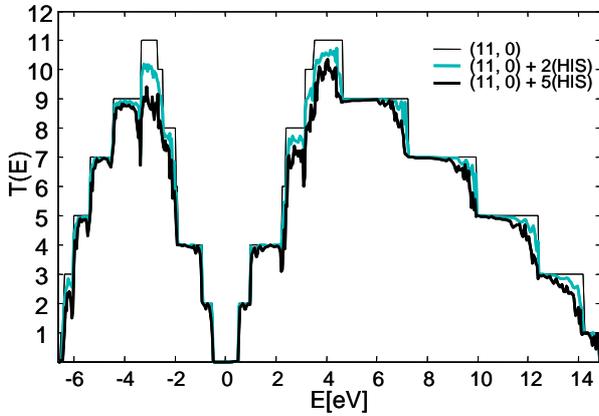

Figure 7. Transmission spectrum of the device in Figure 6 for the pristine (11, 0) nanotube and the nanotube plus amino acids.

Transmission spectrum $T(E)$ was then computed using Equations (14-17) and finally (9), where the transmission is simply $T(E) = G(E)/(2e^2/h)$. Figure 7 shows the lowering of the transmission as increasingly many Histidines adsorb at the surface of the tube.

In conclusion we have presented a framework for studying carbon nanotube-based conductance sensors. We have developed a method that allows the parameterization of tight binding-like systems of low symmetry. A very efficient conductance formula was derived for the case of carbon nanotubes and was applied to a toy model to demonstrate that carbon nanotubes can be considered for amino acid sensing.